%% file: main.tex
\documentclass[10pt,twocolumn,letterpaper]{article}

\usepackage[pagenumbers]{cvpr}      

\input{preamble}

\definecolor{cvprblue}{rgb}{0.21,0.49,0.74}
\usepackage[pagebackref,breaklinks,colorlinks,citecolor=cvprblue]{hyperref}

\usepackage{multirow}
\usepackage{graphicx}
\usepackage{xcolor}
\usepackage{soul}
\usepackage{algorithm}
\usepackage{algorithmic}
\usepackage{bbm}
\usepackage{booktabs}

\title{BrainLeaks: On the Privacy-Preserving Properties of Neuromorphic Architectures against Model Inversion Attacks }

\author{Hamed Poursiami\\
George Mason University\\
Fairfax, Virginia, USA\\
{\tt\small hpoursia@gmu.edu}
\and
Ihsen Alouani\\
Queen's University Belfast\\
Belfast, UK\\
{\tt\small i.alouani@qub.ac.uk}
\and
Maryam Parsa\\
George Mason University\\
Fairfax, Virginia, USA\\
{\tt\small mparsa@gmu.edu}
}

\begin{document}
\maketitle
\input{sec/0_abstract}    
\input{sec/1_intro}
\input{sec/2_Preliminaries}

\input{sec/3_methodology}

\input{sec/4_Experiments}

\input{sec/7_conc}

{
    \small
    \bibliographystyle{IEEEtran}
    
    \bibliography{main}
}


\end{document}

%% file: preamble.tex
\usepackage[dvipsnames]{xcolor}

%% file: sec/0_abstract.tex
\begin{abstract}
\noindent
With the mainstream integration of machine learning into security-sensitive domains such as healthcare and finance, concerns about data privacy have intensified. Conventional artificial neural networks (ANNs) have been found vulnerable to several attacks that can leak sensitive data. Particularly, model inversion (MI) attacks enable the reconstruction of data samples that have been used to train the model. Neuromorphic architectures have emerged as a paradigm shift in neural computing, enabling asynchronous and energy-efficient computation. However, little to no existing work has investigated the privacy of neuromorphic architectures against model inversion. Our study is motivated by the intuition that the non-differentiable aspect of spiking neural networks (SNNs) might result in inherent privacy-preserving properties, especially against gradient-based attacks. To investigate this hypothesis, we propose a thorough exploration of SNNs' privacy-preserving capabilities. Specifically, we develop novel inversion attack strategies that are comprehensively designed to target SNNs, offering a comparative analysis with their conventional ANN counterparts. Our experiments, conducted on diverse event-based and static datasets, demonstrate the effectiveness of the proposed attack strategies and therefore questions the assumption of inherent privacy-preserving in neuromorphic architectures.

\end{abstract}

%% file: sec/1_intro.tex
\section{Introduction}
\label{sec:intro}

The widespread adoption of neural networks in various applications, such as facial recognition \cite{taigman2014deepface} and healthcare \cite{hc}, has raised significant privacy concerns. These neural networks are trained on large datasets that may contain sensitive information. While the original datasets are kept confidential, the trained models are often released publicly assuming they do not leak data. However, studies have shown that adversaries are able to infer information on the training dataset only with having access to the model \cite{subbanna2021analysis, MIA_shokri,E_mia_Shokri}.

One notable privacy threat is Model Inversion (MI) attacks\cite{fredrikson2014privacy, NEURIPS2021_MI, cvpr21_MI}, wherein an adversary attempts to exploit the model's outputs to reconstruct the training data. For instance, in facial recognition systems, malicious users may use MI attacks to partially reconstruct sensitive face images used in the training process~\cite{fredrikson2015model} .
One of the first MI attacks on neural networks was introduced in \cite{fredrikson2015model}, where a gradient descent algorithm was used to find a reconstructed input most likely to be classified as the target label. Several follow-up approaches have been later proposed to enhance the  efficiency of this attack for several ANN models \cite{zhang2020secret,wang2021variational,chen2021knowledge,nguyen2023re}, highlighting the critical aspect of these attacks in breaching data privacy and confidentiality. 

Neuromorphic architectures have promising properties, especially in terms of ultra-low power consumption, which accelerates their integration into the wide ML landscape with platforms like Intel's Loihi \cite{loihi}. Besides, in contrast with ANNs, SNNs' architecture involves non-differentiable operators that operate on discrete spike-based events (more details in Section \ref{sec:preliminaries}). Therefore, towards training these models, several approaches have been proposed to use surrogate gradients in the backward pass \cite{neftci2019surrogate}.  Since the efficiency of MI attacks against ML models is closely related to the capacity of estimating the gradient over the input, this suggests the \textit{hypothesis of potential inherent privacy properties of SNNs}. 

Although there was a raising interest in the security and privacy issues in the ML community, there is little research work that focuses on the privacy-preserving aspect of neuromorphic architectures. For examples, \cite{dpsnn} and \cite{DP_snn} investigated the implementation of differential privacy to protect SNNs. Other effort has been proposed for the deployment of homomorphic encryption for privacy-preserving inference, taking advantage from the sparsity of the spiking data \cite{FHE1,FHE2}. In \cite{panda}, the authors investigate the leakage during the conversion ANN-SNN. However, none of the existing work investigated the inherent properties of SNNs in terms of privacy-preserving.

In this work, we comprehensively investigate the privacy properties of SNNs from MI perspective. The challenge of MI in the spiking domain is that the estimated gradients over the input cannot be directly applied to update discrete binary input spikes, and therefore new spike-compatible approaches need to be investigated.  To address this challenge, we propose two different spike-compatible attacks to infer private spike-train samples from SNN models. The first, \textbf{BrainLeaks-v1} is a spike-based model inversion attack that projects the surrogate gradient into the spiking domain to generate spike-compatible inverted input. The second, \textbf{BrainLeaks-v2} relies the representation of spiking data as Bernoulli distributions and aims to estimate the parameters of these distributions. Our findings indicate that while SNNs generally exhibit greater resilience compared to ANNs, they still suffer from substantial privacy risks. Our key contributions are summarized as follows:

\vspace{12pt}
 \begin{itemize}
     \item We propose an investigation of privacy-preserving capabilities of SNNs. Specifically, we thoroughly study SNNs' privacy under model inversion attacks, comparatively with ANNs.


     \item We first propose \textbf{BrainLeak-v1}, as a spike-based model inversion attack that projects the continuous surrogate gradients obtained during backpropagation into the discrete spiking domain, enabling the generation of spike-compatible inverted inputs.

    \item We propose \textbf{BrainLeak-v2}, which models spike trains as Bernoulli distributions and reframes the MI problem into the estimation of the distribution parameters. \textbf{BrainLeaks-v2} succeeds in extracting private information from SNNs on both static and dynamic spiking datasets with an efficiency that is comparable to the ANNs with input in the image domain.
    
 \end{itemize}

\noindent
To our knowledge, this is the first rigorous investigation of model inversion vulnerabilities in SNNs. The proposed attacks shed light on the privacy risks and advantages of neuromorphic architectures, motivating further research into data privacy for this emerging neural computing paradigm. We make our code open source for reproducibility.


%% file: sec/2_Preliminaries.tex
\section{Preliminaries: Spiking Neural Networks}
\label{sec:preliminaries}

SNNs are brain-inspired neural networks that more closely emulate the temporal dynamics of biological neurons compared to traditional models. Unlike conventional artificial neurons that utilize continuous activation values, spiking neurons communicate through discrete spikes over time \cite{schuman2022opportunities}. SNN neurons accumulate incoming spikes and generate an output spike when their membrane potential reaches a threshold. Several spiking neuron models have been proposed to simulate spiking behavior, with the leaky integrate-and-fire (LIF) model being one of the most prevalent \cite{eshraghian2023training}. The dynamics of a single LIF neuron is mathematically represented by the following discretized equation:
\begin{equation}
    \nu [n] = \alpha \cdot \nu [n-1] + \sum_{k}^{}\omega_k \cdot I _{\text{k}}[n] - O[n-1] \cdot \theta
\end{equation}
Where $n$ represents discretized time steps, $\nu$ is the membrane potential, and $\alpha$ is the leakage decay factor. $I _{\text{k}}$ and $\omega_k$ denote input spikes arriving from presynaptic neuron $k$ and their corresponding synaptic weights, respectively. $O[n]$ is the activation function defined as:
\begin{equation}
O[n]=\left\{\begin{matrix}
1 & ,if \ \nu[n]>\theta\\ 
0 & ,otherwise.
\end{matrix}\right.
\label{eq:activation}
\end{equation}
This neuron model generates the output spikes when the membrane potential surpasses the firing threshold $\theta$, and incorporates a soft-reset mechanism, where the threshold is subtracted from the membrane potential upon the generation of a spike.

In SNNs, the input data must be compatible with the neuron's spiking communication method. This spiking input often comes directly from event-based sensors like dynamic vision sensors (DVS) cameras, which inherently produce spike. Alternatively, inputs from static data are transformed into spikes using encoding schemes such as rate encoding. In this approach, input values are converted into firing rate \cite{gerstner2014neuronal}.

One possible approach to train SNNs is to use methods similar to those in sequential models such as backpropagation through time (BPTT) \cite{bellec2018long}. However, BPTT faces challenges due to the non-differentiable nature of spike events. An effective solution to this issue is to substitute the non-differentiable Heaviside activation function with a continuous and smooth surrogate function during the backward pass.

%% file: sec/3_methodology.tex
\section{MI Attacks for \mbox{Neuromorphic} Architectures}
\label{sec:method}

\subsection{Problem Formulation}
In MI attacks, an adversary exploits access to a target model, denoted as $M$, which is trained on confidential data $D$. The goal is to reconstruct an input, $\hat{x}$, so it is classified as a desired label $y$. 
This reconstruction is an optimisation that minimizes an identity loss, $L_{\text{id}}$, to increase the likelihood of $\hat{x}$ being classified as $y$. $L_{\text{id}}=1-M_y(\hat{x})$, where $M_y(\hat{x})$ represents the posterior probability of $y$ as predicted by $M$ for $\hat{x}$.

To minimize this loss via gradient descent, it is necessary to compute the gradient of the loss with respect to the inputs, represented as $\nabla_{\hat{x}} L$. This gradient is typically derived through backpropagation. However, the activation functions of LIF neurons (as described in equation~\ref{eq:activation}) result in Delta Dirac gradient functions, which hinder backward gradient flow. To overcome this issue, we utilize a surrogate function in the backward pass. 
In this work, we consider the commonly used fast sigmoid function as the surrogate function \cite{zenke2021brain}. 

The challenge of MI in the spiking domain is that the obtained gradients cannot be directly applied to update discrete binary input spikes. To address this incompatibility, we develop two inversion methodologies tailored for SNNs:

\subsection{BrainLeaks-v1}

\begin{figure}[tb]
    \centering
    \includegraphics[width=\linewidth]{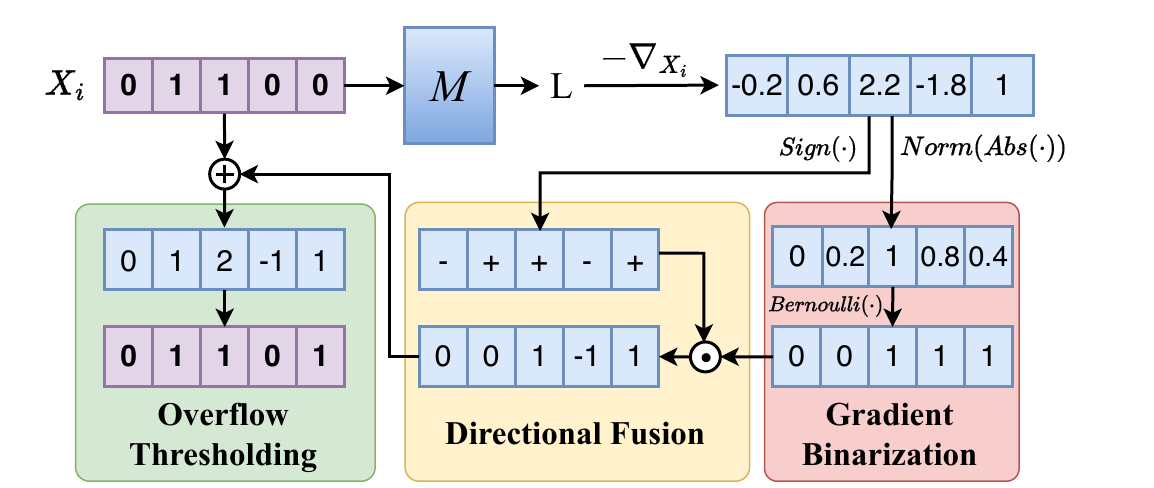}
    \caption{Illustrative overview of an iteration of \textbf{BrainLeaks-v1} to update the input.} 
    \label{fig:G2S}
\end{figure}

In this approach, we leverage surrogate functions during backpropagation, and derive gradients of the loss with respect to the spike-based inputs. To address the incompatibility between the continuous gradients and the discrete spikes, \textbf{BrainLeaks-v1} is inspired from the G2S conversion strategy proposed by \cite{liang2021exploring} in the context of generating adversarial attacks. Specifically, we quantize the surrogate gradients to preserve the spike format of the inputs. This process is outlined in three key steps illustrated in Figure ~\ref{fig:G2S} and summarized below:\\

\noindent \textit{i) Gradients Binarization}: In this initial phase, the absolute values of the gradients are normalized to the range $[0,1]$. Subsequently, we derive binerized gradients $G_{i}$ as follows: 

\begin{equation}
G_{i} = Bernoulli \left ( Norm\left | \nabla_{X_i} \right | \right )
\end{equation}

\noindent Here, $Bernoulli(p)$ refers to the Bernoulli distribution, which produces binary outputs based on the probability $p$, and $Norm$ identifies as normalization by the maximum value of the vector.\\

\noindent \textit{ii) Directional Fusion}: The original gradient signs are combined with the sampled binary mask. This incorporates the directionality of the gradients onto the mask, forming ternary discrete gradients of {-1,0,1}.\\

\noindent \textit{iii) Overflow Thresholding}: The ternary gradients are used to directly update the input spikes. This update might potentially change spike values to fall outside the valid set of ${0, 1}$, leading to invalid values of -1 or 2. Therefore, we clip the updated spikes, converting out-of-bounds values (-1 and 2 to 0 and 1, respectively) and keeping valid spike values (0 and 1) unchanged.\\

The primary challenge of this approach, particularly in its applicability to Inversion Attacks, lies in the normalization step. This step impedes the convergence of the gradient mask, causing the update process to potentially continue indefinitely. Consequently, the attack becomes overly sensitive to hyperparameters, such as the number of iterations and the initialization settings. Moreover, our experiments indicate that this method significantly struggles with the increasing dimensionality and complexity of input data, especially for neuromorphic event-based data such as DVS. 

\subsection{BrainLeaks-v2}

\begin{figure*}[tb]
    \centering
    \includegraphics[width=0.7\linewidth]{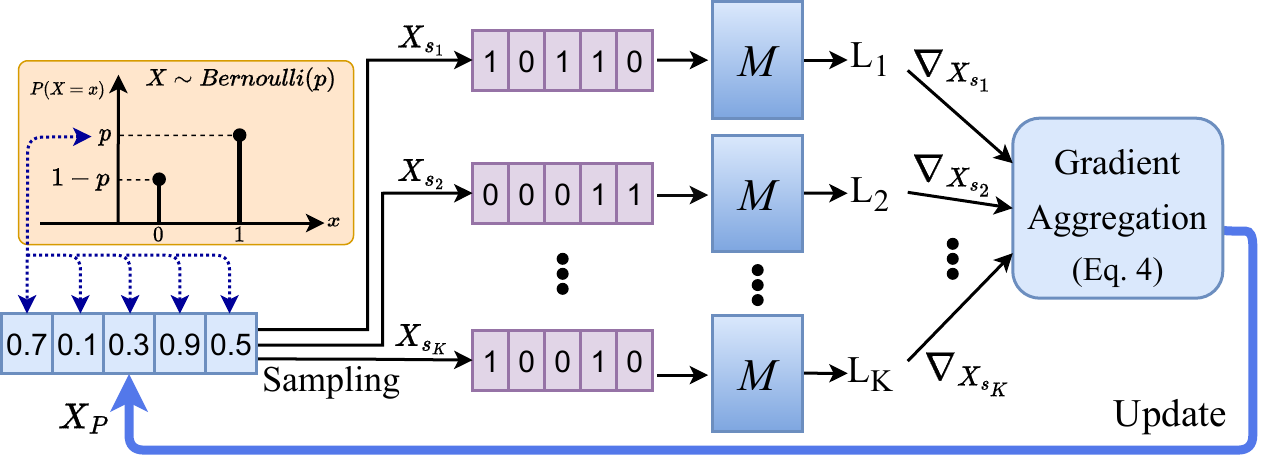}
    \caption{A high-level overview of the \textbf{BrainLeaks-v2} inversion process for a single-feature input spike train}
    \label{fig:BL}
\end{figure*}

In this approach we formulate the problem differently to take into account the inherent properties of the spiking data. We observe that the same (static) input can be represented in the spiking domain with different approaches, even for the same encoding (e.g. rate encoding). Therefore, instead of optimizing for a specific input (supposedly deterministic), we model the input that we want to retrieve as a Bernoulli distribution. Hence, rather than identifying deterministic input spikes that minimize the loss, our approach seeks the expected Bernoulli distribution's parameters  that minimize the expected loss. 
 An overview of this approach is illustrated in Figure~\ref{fig:BL}.

Since the expected value of a Bernoulli random variable corresponds to its probability of success, this method allows us to modify the associated continuous parameters, i.e. probabilities of success. These adjusted probabilities can then be employed to generate reconstructed sample inputs. 

Formally, consider $X_p$ as a tensor that holds values ranging from  $0$ to $1$, matching the spatiotemporal dimensions of the input to the target model. During each iteration of BrainLeaks-v2, the values of $X_p$ act as Bernoulli parameters. These parameters are used to generate a spike-based input $X_s$, which is subsequently fed into the model to calculate the loss. The gradients with respect to the input are then obtained through backpropagation, using surrogate functions. These gradients are essential for updating the values in 
This process, however, introduces a certain level of noise into the optimization trajectory, which can adversely affect the convergence of the optimization algorithm. To mitigate this, we implement an additional strategy inspired by natural evolution strategies (NES), as suggested by~\cite{xu2023sparse}. Specifically, instead of generating a single sample from $X_p$ in each iteration, we produce a population of $K$ samples.

For each sample, we calculate the corresponding loss and gradient. These gradient vectors are then aggregated through a weighted summation, where the weights are determined by the respective losses. This results in a lower variance gradient estimation, which can be expressed as:

\begin{equation}
\label{eq:Gradient Estimation}
    \nabla_{X_{p}}L \approx  \frac{\sum_{i=1}^{K}e^{-L_{i}} \cdot \nabla_{X_{s_{i}}}L_{i} }{\sum_{i=1}^{K}e^{-L_{i}}}
\end{equation}

By implementing this NES-based strategy, we are able to integrate the advantages of both stochastic and deterministic gradient information during the optimization. This approach helps to smooth out the noise and improve the convergence of the algorithm.

A key advantage of SNNs is their ability to handle sparse input formats. We capitalize on this feature by introducing a regularization mechanism in the inversion process to avoid solutions with excessive spikes that do not reflect the expected sparsity. Specifically, we add a penalty term to the cost function to regulate sparsity based on the ratio of spikes in the input:

\begin{equation}
P_{\xi} = \xi \cdot \frac{\text{\# of Spikes}}{\text{Total \# of Voxels}}
\end{equation}
In this equation, $\xi$ represents the strength of the sparsity regularization. A higher $\xi$ value penalizes solutions with more spikes, directing the optimization towards reconstructing inputs with sparse spike patterns. This approach aims to identify the crucial spikes that contribute to an increased confidence in the output. 

It is important to note that we employ a two-stage clamping and scaling approach to maintain the values of $X$ between $0$ and $1$.  In this approach, negative values are clamped to $0$, while positive values exceeding 1 are scaled to ensure the preservation of relative differences between elements. Furthermore, to enhance stability and reduce oscillations in the trajectory, we integrate adaptive learning rates, inspired by RMSProp, along with momentum. The complete procedure is detailed in Algorithm \ref{alg:1}.

\begin{algorithm}[t]
\caption{BrainLeaks-v2}
\begin{algorithmic}[1]
 \renewcommand{\algorithmicrequire}{\textbf{Input:}}
 \renewcommand{\algorithmicensure}{\textbf{Initialize:}}
 
 \REQUIRE Target Model ($M$), Target Class ($y$), NES Population Size ($K$), Sparsity Penalty Strength ($\xi$), RMSProp Decay Rate ($\rho$), Momentum Coefficient ($\beta$), Learning Rate ($\eta$)
 \ENSURE Spiking Distribution Parameters ($X_p$),\\ RMSProp Accumulation Variable ($r$),\\ Momentum Accumulation Variable ($\nu$)
  \WHILE{stopping criterion not met}

    \FOR {i = 1 to K}
  
     \STATE $X_{s_i} = Bernoulli(X_{p})$ \COMMENT{Bernoulli Sampling}
     \STATE $L_{i}=1-M_y(X_{s_i}) + P_{\xi}$ \COMMENT{Loss Calculation}  
    \ENDFOR
  \STATE $\nabla_{X_{p}} \approx  \frac{\sum_{i=1}^{K}e^{-L_{i}} \cdot \nabla_{X_{s_{i}}} }{\sum_{i=1}^{K}e^{-L_{i}}}$ \COMMENT{NES Gradient Estimation}

  \STATE $r \leftarrow \rho r + (1-\rho) \nabla_{X_{p}}^{2}$ \COMMENT{RMSProp accumulation}
  \STATE $\eta \leftarrow \frac{\eta}{\sqrt{r}}$ \COMMENT{Adaptive Learning Rate}

  \STATE $\nu \leftarrow \beta \nu + \nabla_{X_{p}}$ \COMMENT{Momentum accumulation}

  \STATE $X_{p} \leftarrow X_{p} - \eta \nu$ \COMMENT{Optimization Update}

  \STATE $Clamp(X_{p}, 0 ,1)$ \COMMENT{Clamp within [0,1]}
 
 \ENDWHILE
 \RETURN $X_{p}$ 
\end{algorithmic} 
\label{alg:1}
\end{algorithm}

This algorithm returns the estimated Bernoulli parameters of the inferred spiking input distribution, rather than deterministic reconstructed spikes. The model inversion attack is completed by sampling from this distribution to generate probable spike patterns that potentially leak private information.

%% file: sec/4_Experiments.tex
\section{Experiments}
\label{sec:experiments}

\subsection{Setup}
To shed light on the inherent privacy-preserving capabilities of SNNs, we compare the inversion quality with that of conventional ANNs trained on the same static datasets and architectures. For the ANN models, the MI-FACE method, proposed by \cite{fredrikson2015model}, serves as our baseline inversion approach. We evaluate the efficacy of BrainLeaks on SNN models trained with both static image and neuromorphic event-based datasets.\\
\noindent\textbf{Datasets.}
We evaluate our method across three tasks: face recognition, digit classification, and event-based gesture recognition. For face recognition, we use the AT\&T Face Database, which includes 400 grayscale images of faces from 40 unique subjects \cite{samaria1994orl}. For digit classification, we utilize the MNIST dataset \cite{6296535} and its neuromorphic counterpart, N-MNIST. N-MNIST comprises recordings of MNIST digits captured using an Asynchronous Time-based Image Sensor (ATIS) \cite{orchard2015converting}. For the neuromorphic gesture recognition, we employ the IBM DvsGesture Dataset. This dataset features 1,342 samples from 10 hand gesture classes, recorded as event streams using a DVS camera \cite{amir2017low}.\\
\noindent\textbf{Models.} To ensure a fair and consistent comparison between ANN and SNN, we use the same network topology as the ANN baseline attack presented in \cite{fredrikson2015model}. This topology consists of a fully connected network with a single layer of 3000 hidden neurons. However, instead of sigmoid activations, in the SNNs we employ LIF neurons with a leakage rate of 0.7. The network is trained using backpropagation through time (BPTT), utilizing a fast-sigmoid function with a slope of 40 for the surrogate gradient calculation. The loss function incorporates cumulative softmax cross-entropy applied to the membrane potentials of the output layer across all time steps.

For data preprocessing, static image data are converted into 25-step spiking representations using a rate-encoding scheme, ensuring compatibility with the neuromorphic architecture. For the event-based datasets, N-MNIST and IBM DvsGestures, spike events are aggregated into discrete time windows, with the constraint that no voxel contains more than one event. Additionally, to align with fully connected network requirements, we only retain events of a single polarity. All experiments in this study were conducted using PyTorch framework, with SNNtorch \cite{eshraghian2023training} allowing incorporation of spiking neuron models. Table \ref{tab:models_targ} summarizes the validation accuracies of the target models on static and neuromorphic datasets.

\begin{table}[b]
\resizebox{\columnwidth}{!}{%
\begin{tabular}{@{}clccccclcl@{}}
\toprule
\multicolumn{2}{c}{\textbf{Dataset}}   & \multicolumn{2}{c}{\textbf{MNIST}} & \multicolumn{2}{c}{\textbf{AT\&T Face}} & \multicolumn{2}{c}{\textbf{N-MNIST}} & \multicolumn{2}{c}{\textbf{IBM Gestures}} \\
\multicolumn{2}{c}{{Type}}      & ANN              & SNN             & ANN                & SNN                & \multicolumn{2}{c}{SNN}              & \multicolumn{2}{c}{SNN}                   \\ \midrule
\multicolumn{2}{c}{\textbf{Val. Acc.}} & 97.96\%& 96.62\%& 95.00\%& 90.00\%& \multicolumn{2}{c}{96.88\%}          & \multicolumn{2}{c}{78.34\%}               \\ \bottomrule
\end{tabular}%
}
\caption{Accuracy of target models on validation data}
\label{tab:models_targ}
\end{table}

\noindent\textbf{Evaluation Metrics.}
We assess the performance of the inversion attack in extracting private information using two approaches: qualitative and quantitative. \mbox{\ul{\textit{Qualitatively}}},  we visually inspect the reconstructed samples, employing time-averaged (rate-decoded) versions of spiking data. \mbox{\ul{\textit{Quantitatively}}}, we
train an \textit{``evaluation classifier"}, which is distinct from the target model, on the same private dataset. This classifier is a proxy for human inspection and is intended to determine whether the reconstructed samples contain private information. Table~\ref{tab:models_eval} summarizes the validation accuracy of \textit{``evaluation classifiers"} on different datasets.

\textit{\ul{Attack Performance Metrics:}} We utilize four different metrics to evaluate the performance of attack. 1. Attack Accuracy, 2. Top-3 Attack Accuracy, 3. Average Confidence, and 4. Distinctive Attack Accuracy (DAA). 

The \textbf{\textit{Attack Accuracy} metric} quantifies the extent of private information leakage based on the prediction accuracy of the evaluation classifier on the reconstructed samples. For each dataset, we utilized convolutional neural networks (CNNs) with a uniform architecture, consisting of two convolutional layers. The first layer contains 12 filters, followed by a second layer with 24 filters. Each layer has 5x5 kernels and is followed by 2x2 max-pooling. In scenarios involving spiking data, whether it's event-based or rate-encoded static, we adopted network by replacing the ReLU activations with LIF neurons.

\begin{table}[b]
\resizebox{\columnwidth}{!}{%
\begin{tabular}{@{}clccccclcl@{}}
\toprule
\multicolumn{2}{c}{\textbf{Dataset}} & \multicolumn{2}{c}{\textbf{MNIST}} & \multicolumn{2}{c}{\textbf{AT\&T Face}} & \multicolumn{2}{c}{\textbf{N-MNIST}} & \multicolumn{2}{c}{\textbf{IBM Gestures}} \\
\multicolumn{2}{c}{{Type}}    & ANN              & SNN             & ANN               & SNN                 & \multicolumn{2}{c}{SNN}              & \multicolumn{2}{c}{SNN}                   \\ \midrule
\multicolumn{2}{c}{\textbf{Val. Acc.}}        & 98.69\%& 98.43\%& 95.00\%& 97.50\%& \multicolumn{2}{c}{96.59\%}          & \multicolumn{2}{c}{87.92\%}               \\ \bottomrule
\end{tabular}%
}
\caption{Accuracy of evaluation classifiers on validation data}
\label{tab:models_eval}
\end{table}

The \textbf{\textit{Average Confidence}} metric serves as a complement to attack accuracy, providing a more detailed insight into the predictor’s certainty regarding the reconstructed inputs. To calculate this metric, each reconstructed sample is passed through the evaluation classifier. Then, we average the confidence scores that correspond to the correct ground truth labels for these samples.

Beyond the standard attack accuracy metrics, we introduce a novel metric called \textbf{\textit{Distinctive Attack Accuracy (DAA)}}. This metric offers an alternative perspective for evaluation. Unlike traditional attack accuracy, which measures the proportion of individual reconstructed samples correctly predicted by the evaluation classifier, DAA evaluates the entire set of reconstructed samples  collectively across all classes. Specifically, for a given batch of reconstructed inputs that covers all class labels, DAA identifies the reconstructed sample with the highest confidence score for each specific class. It then verifies whether this sample correctly matches with the original attacked class label.

\subsection{Results}

\begin{table*}[ht!]
\resizebox{\textwidth}{!}{%
\begin{tabular}{@{}clcccccccc@{}}
\toprule
\multicolumn{2}{c}{\textbf{Dataset}}          & \multicolumn{2}{c}{\textbf{AT\&T Face}}   & \multicolumn{2}{c}{\textbf{MNIST}}       & \multicolumn{2}{c}{\textbf{N-MNIST}}      & \multicolumn{2}{c}{\textbf{IBM Gestures}} \\
\multicolumn{2}{c|}{\textbf{Method}}          & BL-v1   & \multicolumn{1}{c|}{BL-v2}   & BL-v1 & \multicolumn{1}{c|}{BL-v2}    & BL-v1   & \multicolumn{1}{c|}{BL-v2}   & BL-v1              & BL-v2             \\ \midrule
\multicolumn{2}{c|}{\textbf{Attack Acc.}}     & 37.50& \multicolumn{1}{c|}{71.27 $\pm$ 2.25} & 100.00     & \multicolumn{1}{c|}{100.00} & 20.00& \multicolumn{1}{c|}{46.12 $\pm$ 6.56} & 30.00            & 58.69 $\pm$ 3.37\\
\multicolumn{2}{c|}{\textbf{Top-3 Att. Acc.}} & 62.50 & \multicolumn{1}{c|}{89.28 $\pm$ 1.87} & 100.00     & \multicolumn{1}{c|}{100.00} & 30.00 & \multicolumn{1}{c|}{91.12 $\pm$ 5.43} & 50.00           & 81.41 $\pm$ 3.59\\
\multicolumn{2}{c|}{\textbf{Avg. Conf.}}      & 34.09        & \multicolumn{1}{c|}{42.04} & 93.11& \multicolumn{1}{c|}{94.59}  & 16.29& \multicolumn{1}{c|}{50.00} &     31.19& 54.24\\
\multicolumn{2}{c|}{\textbf{DAA}}             & 90.00& \multicolumn{1}{c|}{100.00}    & 100.00& \multicolumn{1}{c|}{100.00}    & 50.00& \multicolumn{1}{c|}{100.00}    & 100.00& 100.00\\ \bottomrule
\end{tabular}%
}
\caption{Quantitative results of \textbf{BrainLeaks-v1} (BL-v1) and \textbf{BrainLeaks-v2} (BL-v2) attacks on SNN models for all datasets.}
\label{tab:snn_attacks}
\end{table*}

\begin{figure*}[ht!]
    \centering
    \includegraphics[width=1\linewidth]{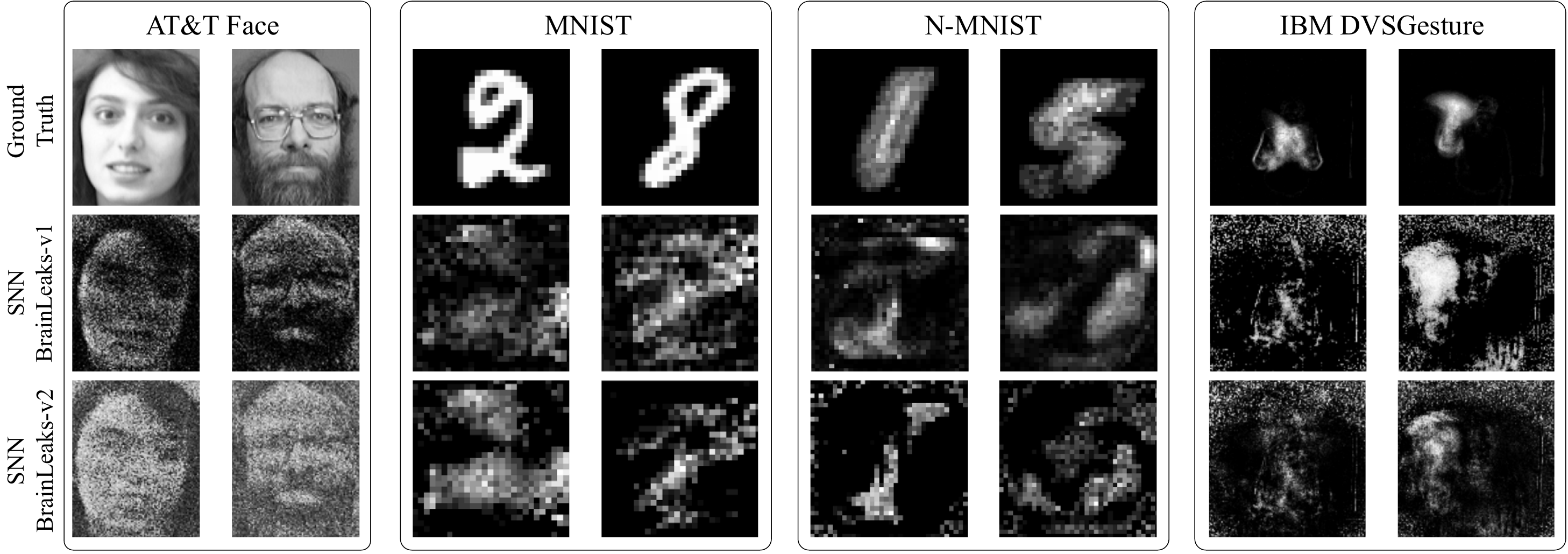}
    \caption{Qualitative results of \textbf{BrainLeaks-v1} and \textbf{BrainLeaks-v2} attacks on SNN models for all datasets.}
    \label{fig:snn}
\end{figure*}

\noindent\textbf{Spiking-only Attacks Evaluation.}
Table~\ref{tab:snn_attacks} presents the results of attack on SNNs using our two proposed methods: BrainLeaks-v1 and BrainLeaks-v2. The BrainLeaks-v1 approach encounters difficulties with more complex datasets, particularly those with higher temporal dimensions. However, in most cases, both attacks register high scores in Distinctive Attack Accuracy (DAA), highlighting the privacy vulnerabilities inherent in SNNs. Notably, although some reconstructed inputs do not accurately match their ground truth classes, the evaluator still identifies specific samples within class that strongly retain characteristics of the original data. This indicates that DAA effectively uncovers partial attack efficacy that might be overlooked when evaluating samples individually. The combination of high DAA with low attack accuracy suggests that while the inversion attack successfully extracts distinctive ``corner cases" associated with each class, it faces challenges in consistently reconstructing the full data distribution.

It is important to highlight that in all the attacks conducted across various datasets, the confidence score exceeded 0.99 in the target classifier, with the notable exception of the BrainLeaks-v1 conversion-based method on the N-MNIST dataset. The BrainLeaks-v1 attack encountered significant convergence difficulties during optimization within this dataset, particularly failing to find inputs that could maximize the confidence scores for the digits $4$ and $9$. This challenge is reflected in the lowest DAA score among all attacks, indicating a strong limitation of the BrainLeaks-v1 method in the context of model inversion attacks. This limitation is also evident in Figure~\ref{fig:snn}, which presents qualitative visualizations of all SNN attacks. In contrast, the BrainLeaks-v2 method demonstrated promising attacks across all datasets, successfully extracting sensitive data. However, the transition from static data to neuromorphic event-based data presents additional complexities for the execution of attacks. As a result, the attack accuracy scores for neuromorphic data are not as high as those achieved with static ones.

It’s also worth mentioning that, although the rate-decoded visualizations of the neuromorphic data display some level of fidelity, the DVS video visualizations show minimal time-varying changes, which indicates that the temporal dependencies are largely preserved, even during successful BrainLeaks attacks. This emphasizes the privacy-preserving characteristics of SNNs.

Another interesting observation is the higher DAA score for BrainLeaks-v1 when attacking the IBM Gesture dataset compared to BrainLeaks-v2, although its other attack metrics are significantly lower. This suggests that the DAA score might be indicative of MI overfitting: a high DAA combined with low attack accuracy implies that the reconstructed samples mostly incorporate global characteristics of the data distribution (the prior distribution rather than the class-conditional ones) but are slightly modified to fit the class characteristics of the target model. These reconstructed samples usually look very similar. For instance, in Figure ~\ref{fig:snn}, the reconstructed sample of digit “5” for N-MNIST closely resembles digit “0”, with minor differences that have caused it to be overfitted, resulting in high confidence for digit “5” when processed by the target model. This exemplifies one of the most significant challenges with BrainLeaks-v1 and will be discussed further when investigating the transferability of our attacks.

\paragraph{Comparison between ANN and SNN.}
To assess the relative privacy vulnerabilities of SNNs and ANNs, we conducted MI attacks on the models trained with the same static datasets - MNIST and the AT\&T Face Database. The results of these attacks, as presented in Table ~\ref{tab:snnann_attacks}, challenge our initial assumption that SNNs might offer greater privacy protection than ANNs.

\begin{table}[b]
\resizebox{\columnwidth}{!}{%
\begin{tabular}{@{}clcccc@{}}
\toprule
\multicolumn{2}{c}{\textbf{Dataset}} & \multicolumn{2}{c}{\textbf{MNIST}}                 & \multicolumn{2}{c}{\textbf{AT\&T Face}} \\ \midrule
\multicolumn{2}{c|}{Type}            & \textbf{ANN}     & \multicolumn{1}{c|}{\textbf{SNN}}        & \textbf{ANN}         & \textbf{SNN}              \\
\multicolumn{2}{c|}{Method}          & MI-Face \cite{fredrikson2015model} & \multicolumn{1}{c|}{BL-v2} & MI-Face \cite{fredrikson2015model}     & BL-v2       \\ \midrule
\multicolumn{2}{c|}{\textbf{Attack Acc.}}     & 50.00& \multicolumn{1}{c|}{100.00}     & 87.50       & 71.27 $\pm$ 2.25\\
\multicolumn{2}{c|}{\textbf{Top-3 Att. Acc.}} & 90.00   & \multicolumn{1}{c|}{100.00}     & 97.50& 89.28 $\pm$ 1.87\\
\multicolumn{2}{c|}{\textbf{Avg. Conf.}}      & 45.38& \multicolumn{1}{c|}{94.59}      & 85.51& 42.04\\
\multicolumn{2}{c|}{\textbf{DAA}}             & 80.00& \multicolumn{1}{c|}{100.00}        & 95.00& 100.00\\ \bottomrule
\end{tabular}%
}
\caption{Comparison of attack performance between ANN and SNN (using BrainLeaks-v2) on MNIST and AT\&T Face.}
\label{tab:snnann_attacks}
\end{table}

\begin{figure}[ht!]
    \centering
    \includegraphics[width=\columnwidth]{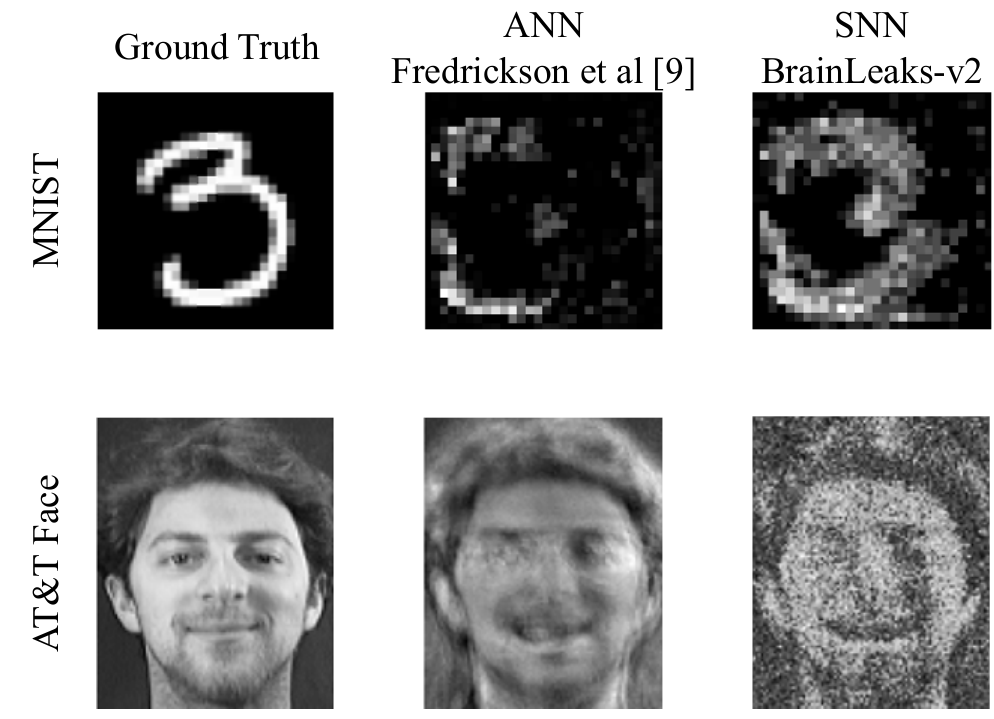}
    \caption{Sample image reconstructions from MNIST and AT\&T Face datasets after MI attacks on ANNs vs SNNs}
    \label{fig:annsnn}
\end{figure}

In the case of the MNIST dataset, the MI attacks on the SNN model was notably successful, extracting all classes with high confidence and outperforming the ANN model. However, in the more complex task of face recognition using the AT\&T face database, the ANNs demonstrated higher attack accuracy and significantly greater confidence levels. This suggests that SNNs may have superior privacy-preserving capabilities in certain contexts only. A qualitative comparison, illustrated in Figure ~\ref{fig:annsnn}, further underscores this dataset-dependent variation. While MNIST reconstructions maintain high fidelity in the SNN model, the AT\&T face samples processed through SNN encoding exhibit a notable loss of recognizable facial features.

The observed dichotomy in the performance of SNNs on different datasets likely arises from the non-reversible quantization inherent in the rate encoder on the SNN input layer. 
For instance, when encoding Face images with 256 intensity levels using 25 time steps, ideally, the reconstructed images would exhibit only 25 distinct intensity levels, assuming no auxiliary or prior information is utilized for reconstruction. While this quantization process may have inadvertently facilitated the inversion of MNIST images, due to their high-contrast simplicity, it also highlights on inherent privacy-preserving characteristic of SNNs. This is particularly evident in the case of the Face dataset, where the quantization significantly impacts the reconstruction fidelity.

%% file: sec/7_conc.tex
\section{Concluding Remarks}
\label{sec:conclusion}
In this work, we investigate the privacy-preserving properties of SNNs, motivated by the potential privacy advantage provided by the non-differentiable functions in \mbox{neuromorphic} models. Particularly, we have explored the privacy risks posed by model inversion attacks to these architectures. We proposed two novel model inversion attack techniques tailored for the spiking domain: BrainLeaks-v1 and BrainLeaks-v2.

Our empirical evaluation, conducted on both static and dynamic spiking datasets, provides new insights into the inherent privacy-preserving capabilities of SNNs. The results demonstrate that SNNs exhibit in most cases slightly superior resilience compared to their ANN counterparts. This resilience can be in part explained by their non-reversible quantization and temporal sparsification, induced by the inherently discrete nature of their computation. Nevertheless, while the BrainLeaks-v1 Conversion-based attack shows high susceptibility to MI overfitting and convergence issues during optimization, our proposed BrainLeaks-v2 algorithm succeeds in extracting private, class-distinctive traits across diverse data types, emphasizing that significant privacy risks still remain, which questions the inherent privacy assumption.

We believe this work represents a first step towards a better understanding of privacy vulnerabilities/properties of neuromorphic architectures. As the adoption of SNNs continues to accelerate across various domains, addressing potential threats will be crucial for developing robust and trustworthy neuromorphic intelligence systems. We hope that our initial analysis will inspire further research into data privacy within the nascent field of spiking neural networks.